\documentclass[preprint,12pt]{elsarticle}

\usepackage{graphics}
\usepackage{amsmath}
\usepackage{amssymb}
\usepackage{graphicx}
\usepackage{epstopdf}
\usepackage{mathtools}

\journal{European Journal of Mechanics - B/Fluids}

\begin{document}

\begin{frontmatter}



\title{ On the intermediate long wave propagation for internal waves in the presence of currents}


\author{Joseph Cullen and Rossen Ivanov \footnote{Corresponding author, Email: rossen.ivanov@tudublin.ie}}

\address{School of Mathematical Sciences, Technological University Dublin, City Campus, Kevin Street, Dublin, D08 NF82, Ireland}

\begin{abstract}
A model for the wave motion of an internal wave in the presence of current in the case of intermediate long wave approximation is studied. The lower layer is considerably deeper, with a higher density than the upper layer. The flat surface approximation is assumed. The fluids are incompressible and inviscid. The model equations are obtained from the Hamiltonian formulation of the dynamics in the presence of a depth-varying current. It is shown that an appropriate scaling leads to the integrable Intermediate Long Wave Equation (ILWE). Two limits of the ILWE leading to the integrable Benjamin-Ono and KdV equations are presented as well. 
\end{abstract}




\end{frontmatter}


\section{Introduction}
\label{intro}

The equatorial region in the Pacific is characterised by a shallow layer of warm and less dense
water over a much deeper layer of cold denser water. The two layers are separated by a sharp
thermocline (where the temperature gradient has a maximum, it is very close to the pycnocline, where the pressure gradient has a maximum) at a depth, depending on the location, but usually at 100 -- 200 m beneath the surface. Both layers are supposed to be homogeneous and their sharp
boundary is the thermocline/pycnocline.
The strong stratification confines the wind-driven currents to a shallow near-surface region. In the Atlantic and Pacific, the westward trade winds induce a westward surface flow at speeds of 25-75 cm/s, while a jet-like current, the Equatorial Undercurrent (EUC), flows below it toward the East (counter to the surface current), attaining speeds of more than 1 m/s at
a depth of nearly 100 m. Since the wind usually moves the top 20 m of
the sea surface, and that the wave speed is usually less than $4$ m/s, the vorticity values are not exceeding by magnitude $0.2 $ rad/s. The values for the lower layer are typically lower.

The EUC flows in a region that is in a very narrow region close to the Equator, it is symmetric about the Equator and extends nearly across the whole length (more than 12000 km) of the Pacific Ocean basin \cite{Iz,FedorovBrown}. With speeds in excess of 1 m/s, the EUC is one of the fastest permanent currents in the world. The flow has nearly two-dimensional character, with small meridional variations, being combinations of longitudinal non-uniform currents and waves, and presenting a significant fluid stratification that results in a pycnocline/thermocline separating two internal layers of practically constant density (see \cite{JMcP}). While at depths in excess of about 240 m there is, essentially, an abyssal layer of still water, the ocean dynamics near the surface is quite complex. In this region the wave motion typically comprises surface gravity waves with amplitudes of 1-2 m and oscillations with an amplitude of 10-20 m at the thermocline (of mean depth between 50 m and 150 m). These internal waves
interact with the underlying current (EUC). The topic of wave-current interactions is a large and important research area within the fluid dynamics and we refer to \cite{CIMT,Constantin_2011,ConstantinEscher,ConstantinEscher2,ConstantinJohnson,Henry,
HM,Jonsson,Peregrine,TelesdaSilvaPeregrine,ThomasKlopman} and the references therein.   

Our goal is to model the wave propagation at the thermocline in the case of intermediate long wave approximation, when the wavelength is comparable to the depth of the lower layer, which in turn is much deeper than the upper layer. We obtain several approximate models and one of them is an equation, integrable by the inverse scattering method. 

The Hamiltonian approach is central to our modelling of internal waves in the presence of current. It originates from Zakharov's paper \cite{Zakharov} for irrotational surface waves over infinitely deep water. The Hamiltonian formalism is often utilized in the study of nonlinear waves in continuous media, see for example the review article \cite{ZK}.  The Lagrangian and the Hamiltonian approaches for surface and internal waves have been developed further in many papers, from which here we mention only
\cite{BenjBridPart1,BenjBridPart2,BenjOlv,CraigGroves1,CraigGuyenneKalisch,CraigGuyenneSul3,Milder,Miles,Ovs}.
Currents with linear profiles and constant vorticity could be included in the Hamiltonian formulation of the wave-current dynamics, recent studies are published in \cite{CompelliIvanov1JNMP,CompelliIvanov3,CompelliIvanov2,CI19,
ConstantinIvanov,CMP19,ConstantinIvanovMartin,IoMa,Iv17}. 

The setup and the governing equations for internal waves are described briefly in Sections \ref{sec:2} and \ref{sec:3}. The Hamiltonian Formulation of the problem is presented in Section \ref{sec:Ham}. The scales leading to the intermediate long wave propagation regime are introduced in Section \ref{sec:5} where the model equations are derived as well. The integrable Intermediate Long Wave Equation (ILWE) is derived in Section \ref{sec:6}. Mathematical details about some integro-differential operators and the integrability of ILWE are given in the Appendix. 

\section{System setup for internal equatorial waves} \label{sec:2}
\begin{figure} 
\centering
\includegraphics[width=0.5\textwidth]{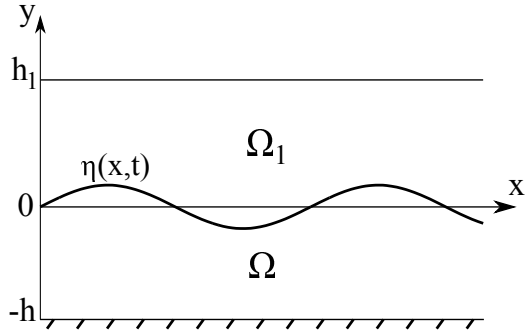}
\caption{System setup. The function $\eta(x,t)$ describes the elevation of the internal wave.}
\label{fig:1}
\end{figure}

The internal wave along the equator is modelled by two layers of water in two dimensions, since the meridional motion is neglected. The layers are separated by a free common interface (the thermocline/pycnocline) as per Figure \ref{fig:1}. 

The system is bounded at the bottom by an impermeable flatbed and is considered as being bounded on the surface by an assumption of absence of surface motion, i.e. rigid lid approximation. The two fluid domains are $\Omega=\{(x, y)\in\mathbb{R}^2: -h< y < \eta(x,t)\}$ and $\Omega_1=\{(x, y)\in\mathbb{R}^2: \eta(x,t)< y < h_1\}. $ The functions and parameters associated with the upper layer will be marked with subscript 1.  Also, subscript $c$ (implying \emph {common interface}) will be used to denote evaluation on the internal wave. Propagation of the internal wave is assumed to be in the positive $x$-direction which is considered to be 'eastward'. The direction of the gravity force is in the negative $y$-axis. 

The function $\eta(x,t)$ describes the elevation of the internal wave with the mean of $\eta$ assumed to be zero, $\int_{\mathbb{R}} \eta(x,t) dx=0.$ Actually it is sufficient $\int_{\mathbb{R}} \eta(x,t) dx=\mathfrak{p}<\infty,$ because the mean value involves division of the length of the corresponding interval, which in this case is  $\int_{\mathbb{R}}  dx =\infty .$ The fluids are incompressible with densities $\rho$ and $\rho_1.$ The stable stratification is given by the immiscibility condition $\rho>\rho_1.$

The stream functions, $\psi$ and $\psi_1$, are related to the velocity fields $\textbf{u}=(u,v)$ and $\textbf{u}_1=(u_1,v_1)$ via the relations
\begin{equation}
\label{definitionofpsi_GenSystem}
\ u=\psi_{y},\quad u_1=\psi_{1,y} ,
     \quad v=-\psi_{x} \quad \mbox{ and } \quad v_1=-\psi_{1,x} 
\end{equation}
due to the incompressibility assumption $\nabla \cdot \textbf{u}=0, $  $\nabla \cdot \textbf{u}_1=0. $

The velocity potentials, $\varphi$ and $\varphi_1$, are introduced such that
\begin{equation}
\label{definitionofpsi_varcurr_GenSystem}
 u = \varphi_{x} +\gamma y, \quad u_1 = \varphi_{1,x} +\gamma_1 y, \quad      v = \varphi_{y} \quad \mbox{ and }  \quad  v_1 = \varphi_{1,y}
\end{equation}
where $\gamma=-v_{x}+u_{y}$ and $\gamma_1=-v_{1,x}+u_{1,y}$ are the constant vorticities. This setup allows for modelling of an undercurrent, such as the Equatorial Undercurrent. A piecewise linear current profile can be represented by the velocity fields of the form (\ref{definitionofpsi_varcurr_GenSystem}), \cite{ConstantinIvanov}
by writing
\begin{equation}
\label{definitionofphi_GenSystem2}
 u =  \widetilde{\varphi}_{x} +\gamma y+\kappa, \quad u_1 =  \widetilde{\varphi}_{1,x} +\gamma_1 y+\kappa_1, \quad
  v=  \widetilde{\varphi}_{y} \quad \mbox{ and } \quad  v_1=  \widetilde{\varphi}_{1,y}
\end{equation}
where  $\kappa$ and  $\kappa_1$ are constants representing the current horizontal velocities at $y=0.$ The wave-only components have been separated out by introducing a tilde notation.  Effectively  the current profiles in $\Omega$  and $\Omega_1$ are $U(y)=\gamma y+\kappa$ and $U_1(y)=\gamma_1 y+\kappa_1$, correspondingly. Here $\kappa$ and $\kappa_1$ are constants, representing the average flow in each respective domain, see 
\cite{ConstantinIvanovMartin,CMP19}, where the situation with a free surface is studied.  

We assume that the functions $\eta(x, t)$, $\widetilde{\varphi}(x, y, t)$ and $\widetilde{\varphi}_1(x, y, t)$ belong to the Schwartz class $\mathcal{S}(\mathbb{R})$ with respect to $x$ (for any $y$ and $t$).  The assumption of course implies that for large absolute values of $x$ the internal wave attenuates, and is vanishing at infinity, and therefore 
\begin{equation}
\label{WaterWaveAssumps}
\lim_{|x|\rightarrow \infty}\eta(x,t)= \lim_{|x|\rightarrow \infty}{\widetilde{\varphi}}(x,y,t) = \lim_{|x|\rightarrow \infty}{\widetilde{\varphi}}_1(x,y,t)=0.
\end{equation}

\section{Governing equations}
\label{sec:3}

The Euler equations for the two layers are
\begin{equation}
\label{Eulers_inert}
 \textbf{u}_{t} +(\textbf{u}.\nabla)\textbf{u}=-\frac{1}{\rho}\nabla p+{\bf g}+\mathbf{F} \quad  \mbox{ and } \quad \textbf{u}_{1,t} +(\textbf{u}_1.\nabla)\textbf{u}_1=-\frac{1}{\rho_1}\nabla p_1+{\bf g}+\mathbf{F_1}
\end{equation}
where 
\begin{equation}
\label{TotalPressure}
\mathbf{F}=2\omega \nabla \psi \quad \mbox{ and }\quad \mathbf{F}_1=2\omega \nabla \psi_1  
\end{equation}
are the Coriolis forces per unit mass at the equator with $\omega$ being the rotational speed of the Earth, ${\bf g}=(0,0,-g)$ is the Earth acceleration, i.e. $g$ is the magnitude of the acceleration due to gravity, $\rho$ and $\rho_1$ (due to the assumption of incompressibility) are the constant densities and $p$ and $p_1$ are the corresponding pressures. The pressure gradients can be expressed as
\begin{align}
\nabla p&=-\rho\nabla\Big(\widetilde{\varphi}_{t} +\frac{1}{2}| \nabla \psi|^2-(\gamma+2\omega)\psi+gy\Big) \notag\\
 \nabla p_1&=-\rho_1\nabla\Big(\widetilde{\varphi}_{1,t} +\frac{1}{2}| \nabla \psi_1|^2-(\gamma_1+2\omega)\psi_1+gy\Big). \notag
\end{align} The dynamic boundary condition at the interface $p=p_1$ gives the Bernoulli equation
\begin{equation}
\label{BernoulliCondition_GenSystem}
\rho\Big((\widetilde{\varphi}_{t})_c +\frac{1}{2}|\nabla \psi|_c^2-(\gamma+2\omega)\chi+g\eta\Big)=\rho_1\Big((\widetilde{\varphi}_{1,t})_c +\frac{1}{2}| \nabla \psi_1|_c^2-(\gamma_1+2\omega)\chi_1+g\eta\Big)
\end{equation}
where the subscript $c$ is a notation for an evaluation at the common interface $y=\eta(x,t),$ $\chi= \psi(x,\eta,t)$ and $\chi_1= \psi_1(x,\eta,t)$. The equation \eqref{BernoulliCondition_GenSystem} gives the time evolution of $\xi:=\rho (\widetilde{\varphi})_c -\rho_1 (\widetilde{\varphi}_1)_c $ which is the {\it momentum} type variable in the Hamiltonian formulation of the problem. The {\it coordinate} variable  is $\eta(x,t)$ and it evolves according to the {\it  kinematic boundary condition} at the interface \cite{Johnson_Book}
\begin{equation}
\label{KBC_GenSystem}
\eta_t=v-u\eta_x=v_1-u_1\eta_x.
\end{equation} This can be expressed in terms of the velocity potentials, using (\ref{definitionofphi_GenSystem2}), as
\begin{equation}
\label{KBC_GenSystem_alt2}
\eta_t=(\widetilde{\varphi}_{y})_c-\big((\widetilde{\varphi}_{x})_c+\gamma \eta+\kappa \big)\eta_x=(\widetilde{\varphi}_{1,y})_c-\big((\widetilde{\varphi}_{1,x})_c+\gamma_1 \eta+\kappa_1 \big)\eta_x.
\end{equation} The kinematic boundary condition at the bottom, requiring that there is no velocity component in the $y$-direction on the flat bed, is given by
\begin{equation}
\label{BottBound_GenSystem}  
   \big(\widetilde{\varphi}(x,-h,t)\big)_y=0 \quad  \mbox{ and } \quad
   \big(\psi(x,-h,t)\big)_x=0
\end{equation}
and, additionally, there is a kinematic boundary condition at the top, requiring that there is no velocity component in the $y$-direction on the surface, given by
\begin{equation}
\label{TopBound_GenSystem}   
     \big(\widetilde{\varphi}_1(x,h_1,t)\big)_y=0 \quad \mbox{ and } \quad
     \big(\psi_1(x,h_1,t)\big)_x=0.
\end{equation}


\section{Hamiltonian formulation}
\label{sec:Ham}
Similar to the case of a single layer with a constant vorticity \cite{NearlyHamiltonian}, the wave-motion on the interface (thermocline/pycnocline) admits a (nearly) Hamiltonian representation. The details could be found for example in \cite{Compelli1Wavemotion,Compelli2MonatshMath,CompelliIvanov3}.
Let us introduce the following notations for the velocity potentials values at the interface  
\begin{equation}
        \left\lbrace
        \begin{array}{lcl}
        \phi:=\tilde{\varphi}(x,\eta(x,t),t) 
        \\
        \phi_1:=\tilde{\varphi}_1(x,\eta(x,t),t) .
        \end{array}
        \right.
\end{equation}
The variable  
\begin{equation} \label{xi}
\xi:=\rho\phi-\rho_1\phi_1
\end{equation}
introduced in \cite{BenjBridPart1,BenjBridPart2} plays the role of a generalised momentum. We assume that $\xi \in \mathcal{S}(\mathbb{R})$ for all $t$. 
The (non-canonical)  equations of motion are 
\begin{equation}
\label{Hsys}
        \left\lbrace
        \begin{array}{lcl}
        \eta_t=\delta_{\xi} H
        \\
        \xi_t=-\delta_{\eta} H+\Gamma  \chi
        \end{array}
        \right.
\end{equation}

\noindent where $H=H(\eta, \xi)$ is the total energy of the system, 
\begin{alignat}{2}
\Gamma:=\rho\gamma-\rho_1\gamma_1+2\omega\big(\rho-\rho_1\big)
\end{alignat}
and
\begin{alignat}{2}
\label{lem2}
\chi(x,t)=- \int_{-\infty}^x\eta_t (x',t)dx'=-\partial_x^{-1}\eta_t
\end{alignat}
is the stream function, evaluated at $y=\eta(x,t).$ 

The dynamics can be formally written in the canonical form 
\begin{equation}
        \left\lbrace
        \begin{array}{lcl}
       \eta_t=\delta_{\zeta} H
        \\
        \zeta_t=-\delta_{\eta} H
        \end{array}
        \right.
\end{equation}

\noindent under the transformation (cf. \cite{Wahlen,Wahlen2})
\begin{alignat}{2}
\label{vartrans}
\xi\rightarrow\zeta=\xi-\frac{\Gamma}{2} \int_{-\infty}^{x} \eta(x',t)\,dx'.
\end{alignat}

The equations \eqref{Hsys}  could be further expressed in terms of the variable  $ \mathfrak{u}=\xi_x  $ for a Hamiltonian written in terms of $\mathfrak{u} $ and $\eta:$
 
 \begin{equation} \label{H_u_eta}
\eta_t =-\Big(\frac{\delta H}{\delta \mathfrak{u}}\Big)_x \quad \text{and} \quad \mathfrak{u}_t=-\Big(\frac{\delta H}{\delta \eta}\Big)_x-\Gamma\eta_t.
\end{equation}

 In order to  obtain the Hamiltonian in terms of variables defined at the interface we need to introduce the so-called Dirichlet-Neumann (DN) operators \cite{CraigGuyenneKalisch,CraigGuyenneSul3}
\begin{equation}
        \left\lbrace
        \begin{array}{lcl}
        G(\eta)\phi=(\tilde{\varphi}_{{\bf{n}}})\sqrt{1+(\eta_x)^2} \mbox{ for $\Omega$}
        \\
        G_1(\eta)\phi_1=(\tilde{\varphi}_{1,{\bf{n}}_1})\sqrt{1+(\eta_x)^2} \mbox{ for $\Omega_1$}
        \end{array}
        \right.
\end{equation}
where $\varphi_{{\bf{n}}}$ and $\varphi_{1,{\bf{n}}_1}$ are the normal derivatives of the velocity potentials, at the interface, and ${{\bf{n}}}$ and ${{\bf{n}}_1}$ are the outward normals to the corresponding domains, with  ${{\bf{n}}}=-{{\bf{n}}_1}$. We need also the operator
\begin{alignat}{2}
\label{B_DEF}
B:=\rho G_1(\eta)+\rho_1 G(\eta).
\end{alignat}

Using the boundary conditions
\begin{alignat}{2}
        \left\lbrace
        \begin{array}{lcl}
        G(\eta)\phi=-\eta_x(\tilde{\varphi}_x)_c+ (\tilde{\varphi}_y)_c = \eta_t+(\gamma\eta+\kappa)\eta_x,
        \\
        G_1(\eta)\phi_1=\eta_x(\tilde{\varphi}_{1,x})_c-(\tilde{\varphi}_{1,y})_c=-\eta_t-(\gamma_1\eta+\kappa_1)\eta_x
        \end{array}
        \right.
\end{alignat}
we obtain $G(\eta)\phi+G_1(\eta)\phi_1=\mu $  where
\begin{alignat}{2}
\mu:=\big((\gamma-\gamma_1)\eta+(\kappa-\kappa_1)\big)\eta_x.
\end{alignat}
With \eqref{xi} we obtain  
\begin{alignat}{2}
\rho_1G(\eta)\phi+\rho G_1(\eta)\phi=\rho_1\mu+ G_1(\eta)\xi.
\end{alignat}
and we can solve for $\phi$ and $\phi_1$:
\begin{alignat}{2}\label{phys}
        \left\lbrace
        \begin{array}{lcl}
        \phi=B^{-1}\big(\rho_1\mu+G_1(\eta)\xi \big)
        \\
        \phi_1=B^{-1}\big(\rho\mu-G(\eta)\xi\big )
        \end{array}
        \right.
\end{alignat}

The expression for the Hamiltonian in terms of $\xi, \eta \in \mathcal{S}(\mathbb{R})$ has the form \cite{CompelliIvanov2}
\begin{align}
\label{Main_Ham}
H(\eta,\xi)=&\frac{1}{2}\int_{\mathbb{R}} \xi G(\eta) B^{-1}G_1(\eta)\xi \,dx
- \frac{1}{2}\rho\rho_1\int_{\mathbb{R}}   \mu   B^{-1}\mu  \,dx- \int_{\mathbb{R}} (\kappa+\gamma\eta)\xi\eta_x \,dx \nonumber \\
&+\rho_1\int_{\mathbb{R}} \mu B^{-1}G(\eta)\xi\,dx +\frac{\rho}{6\gamma}\int_{\mathbb{R}} [(\gamma \eta+\kappa)^3-\kappa^3] dx \nonumber \\&-\frac{\rho_1}{6\gamma_1}\int_{\mathbb{R}} [(\gamma_1 \eta+\kappa_1)^3-\kappa_1^3]dx +\frac{1}{2}g(\rho-\rho_1 )\int_{\mathbb{R }}\eta^2 dx  .
\end{align}

In the special case $\gamma_1=\gamma,$  $\kappa_1=\kappa$ and $\mu=0$ the Hamiltonian acquires the form
\begin{align}
H(\eta,\xi)=&\frac{1}{2}\int_{\mathbb{R}}  \xi G(\eta) B^{-1}G_1(\eta)\xi \,dx -\int_{\mathbb{R}} (\kappa+\gamma\eta)\xi\eta_x \,dx \nonumber \\
&+\frac{\rho-\rho_1}{6\gamma}\int_{\mathbb{R}} [(\gamma \eta+\kappa)^3-\kappa^3]dx +\frac{1}{2}g(\rho-\rho_1 )\int_{\mathbb{R}} \eta^2 dx  \label{HamPlisk}
\end{align}
thus recovering the Hamiltonian determined in \cite{CompelliIvanov3}. When $\gamma_1\neq\gamma$ and $\kappa_1=\kappa=0$ then $\mu=(\gamma-\gamma_1)\eta\eta_x$ and the Hamiltonian becomes
\begin{align}
H(\eta,\xi)=&\frac{1}{2}\int_{\mathbb{R}} \xi G(\eta) B^{-1}G_1(\eta)\xi \,dx
+\rho_1 (\gamma-\gamma_1) \int_{\mathbb{R}}     \eta\eta_x B^{-1}G(\eta)\xi\,dx
\nonumber \\
&- \frac{1}{2}\rho\rho_1(\gamma-\gamma_1)^2 \int_{\mathbb{R}}   \eta\eta_x B^{-1}\eta\eta_x  \,dx -\gamma \int_{\mathbb{R}}   \xi\eta\eta_x \,dx
\nonumber \\
&+\frac{1}{6}(\rho\gamma^2-\rho_1\gamma_1 ^2) \int_{\mathbb{R}}  \eta^3dx +\frac{1}{2}g(\rho-\rho_1 )\int_{\mathbb{R}} \eta^2 dx
\label{HamMonat} \end{align}
which, given the definition of $B$ from (\ref{B_DEF}), recovers the result in \cite{Compelli2MonatshMath} (noting that there is a sign difference is the second and fourth terms due to the different stream function convention used in \cite{Compelli2MonatshMath}).

In the situation with $\kappa_1=\kappa$ which is physically realistic, because the unperturbed currents at the two layers have the same speed at the interface (i.e. absence of a vortex sheet), we have also $\mu=(\gamma-\gamma_1)\eta \eta_x$ and

\begin{align}
\label{Main_approx}
H(\eta,\xi)=&\frac{1}{2}\int_{\mathbb{R}} \xi G(\eta) B^{-1}G_1(\eta)\xi \,dx
- \frac{1}{2}\rho\rho_1 (\gamma -\gamma_1)^2\int_{\mathbb{R}}  
 \eta \eta_x    B^{-1}  \eta \eta _x   \,dx \nonumber \\
 &- \int_{\mathbb{R}} (\kappa+\gamma\eta)\xi\eta_x \,dx
+\rho_1 (\gamma - \gamma_1) \int_{\mathbb{R}} \eta \eta_x B^{-1}G(\eta)\xi\,dx
\nonumber \\ &+\frac{\rho\gamma^2-\rho_1 \gamma_1^2}{6}\int_{\mathbb{R}} \eta^3  dx +\frac{g(\rho-\rho_1)+(\rho \gamma - \rho_1 \gamma_1) \kappa}{2}\int_{\mathbb{R }}\eta^2 dx  .
\end{align} 

The stability of the flow under disturbances of arbitrary wavenumber is assumed. For a detailed study of the stability (which is beyond the scope of our study) we refer to \cite{BV} and the references therein.

In what follows appropriate scales will be introduced and the corresponding expansions of the DN operators will be provided in terms of the small-order scale parameters.

\section{Scales and approximations}
\label{sec:5}
Let us introduce a scale 
\begin{equation}  \varepsilon=\frac{a}{h_1}
\end{equation} where the constant $a$ represents the average amplitude of the waves $\eta(x,t)$ under consideration and $\varepsilon \ll 1$ is a small parameter which will be used to separate the order of the terms in the model.

The Dirichlet-Neumann (DN) operators have the following structure   

\begin{equation} \label{DN}
{G}={G}^{(0)} +{ G}^{(1)}+{G}^{(2)}+ \ldots
\end{equation} 
where ${G}^{(n)}(\eta)$ is an operator, such that  ${G}^{(n)}(\nu \eta)=\nu^n {G}^{(n)}( \eta)$ for any constant $\nu$ i.e. ${G}^{(n)}\sim  \varepsilon^n \sim \eta^n $, since $\eta \sim h_1 \varepsilon $ and similarly for $G_{1}$. The corresponding expansions are \cite{CraigGuyenneKalisch,CraigGuyenneSul3} 



\begin{alignat}{2}
\label{DN_G}
{G}({\eta})&={D}\tanh(h {D})+{D} {\eta} {D} -{D}\tanh(h {D}) {\eta} {D}\tanh(h {D})+\mathcal{O}(\eta^2)\\
\label{DN_G1}
{G}_1({\eta})&={D}\tanh(h_1 {D})-{D} {\eta} {D} +{D}\tanh(h_1 {D}) {\eta} {D}\tanh(h_1  {D})+\mathcal{O}(\eta^2)
\end{alignat} where ${D}=-i\partial/\partial {x}$.

We will study the equations of motion under the additional approximation that the wavelengths $L$ are much bigger than $h_1$, i.e. 
$$\delta=\frac{h_1}{L} \ll 1.$$ Noting that the wave number $k=2\pi/L $ is an eigenvalue or a Fourier multiplier for the operator $D$ (when acting on waves of the form $e^{ikx}$)  we make the following further assumptions about the scales

1. $\delta=\mathcal{O}(\varepsilon);$

2. $h k=\mathcal{O}(1)$ and $h_1k=\mathcal{O}(\delta)$ i.e. $ h_1/h \sim \delta \ll 1$. This corresponds to a deep lower layer;

3. $\xi=\mathcal{O}(1)$. 

4. The physical constants $h_1,$ $\rho,$ $\rho_1,$ $\gamma,$  $\gamma_1$ are $\mathcal{O}(1) .$
 
Since the operator $D$ has an eigenvalue $k$, thus we shall also keep in mind that $hD=\mathcal{O}(1) $  and $h_1D=\mathcal{O}(\delta) .$  Writing the DN operators in the form

\begin{align}
{G}({\eta})=&\frac{1}{ h_1}\left[  (h_1{D})\tanh(h {D})+  (h_1{D}) \frac{{\eta}}{ h_1} ( h_1  {D}) \right.  \nonumber \\
& \phantom{*****}\left. -(h_1{D})\tanh(h {D}) \frac{{\eta}}{  h_1}( h_1 {D})\tanh(h {D})+\ldots \right] \label{DN_G11} \\
{G}_1({\eta})=&\frac{1}{h_1}\left[ (h_1{D})\tanh(h_1 {D})-(h_1 {D}) \frac{{\eta}}{h_1}( h_1{D}) \right. \nonumber \\& \phantom{*****}\left.+(h_1{D})\tanh(h_1 {D})\frac{ {\eta}} {h_1} (h_1 {D})\tanh(h_1 {D})+ \ldots \right].\label{DN_G1}
\end{align}

we can determine explicitly the scale factors 

\begin{align}
{G}({\eta})&=\frac{1}{ h_1}\left[ \delta (h_1\bar{D})\tanh(h \bar{D})+ \delta^3(h_1\bar{D} )\frac{\bar{\eta}}{ h_1} ( h_1  \bar{D}) \right. \nonumber \\
& \phantom{*****}\left. -\delta^3 (h_1\bar{D})\tanh(h \bar{D}) \frac{\bar{\eta}}{  h_1}( h_1 \bar{D})\tanh(h \bar{D})\right] +\mathcal{O}(\delta^4)  \label{DN_G12}\\
{G}_1({\eta})&=\frac{1}{h_1}\left[ \delta(h_1\bar{D})\left(\delta(h_1 \bar{D})-\delta^3\frac{(h_1 \bar{D})^3}{3}\right)  -\delta^3(h_1\bar{D}) \frac{\bar{\eta}}{h_1}( h_1\bar{D}) \right. \nonumber \\ & \phantom{*****} \left.+\delta^5(h_1\bar{D})(h_1 \bar{D})\frac{ \bar{\eta}} {h_1}( h_1\bar{D})(h_1 \bar{D}) \right]+\mathcal{O}(\delta^4) .\label{DN_G14}
\end{align}
where the barred quantities and operators together with $h$ and $h_1$ are assumed to be of order 1. Introducing $\mathfrak{t}_h:=\tanh(hD)$, $\mathcal{D}_1:= h_1 D$ and omitting the bars for convenience we truncate the DN expansions as follows:

\begin{alignat}{2}
\label{DN_G13}
G(\eta)&=\frac{1}{ h_1}\left[ \delta \mathcal{D}_1 \mathfrak{t}_h + \delta^3 \mathcal{D}_1\frac{{\eta}}{ h_1} \mathcal{D}_1 -\delta^3 \mathcal{D}_1 \mathfrak{t}_h \frac{\eta}{  h_1} \mathcal{D}_1 \mathfrak{t}_h   \right] +\mathcal{O}(\delta^4)  \\
\label{DN_G15}
{G}_1({\eta})&=\frac{\delta^2}{h_1}\mathcal{D}_1 \left[ 1-\delta  \frac{{\eta}}{h_1}  -\delta^2 \frac{\mathcal{D}_1^2}{3}\right]\mathcal{D}_1 +\mathcal{O}(\delta^5) .
\end{alignat}
Note that $h$ appears only in the definition of the operator $\mathfrak{t}_h,$  which is of order 1. Since $h_1$ is assumed of order 1, then formally the order of the differentiation $\partial_x$ is $\delta$. Hence the order of the integration measure $dx$ is $1/\delta$.

We see now that the leading order terms in $G$ are $\mathcal{O}(\delta)$ and the leading order terms in $G_1$ are $\delta^2$ hence $G_1 G^{-1} \sim \delta \ll 1$ and hence we can expand as follows: 
\begin{alignat}{2}
\nonumber
GB^{-1}G_1&= G\frac{1}{\rho_1 G + \rho G_1}G_1 = G\frac{1}{\rho_1 (1 + \frac{\rho}{\rho_1}G_1 G^{-1}) G}G_1 \\&=\frac{1}{\rho_1}  G G^{-1}\left[ 1-\frac{\rho}{\rho_1} G_1 G^{-1 }  +    \frac{\rho^2}{\rho_1^2}(G_1 G^{-1})^2  - \ldots\right] G_1 \nonumber \\
 &    =  \frac{1}{\rho_1}  \left[ G_1 -\frac{\rho}{\rho_1} G_1 G^{-1 }G_1 + \frac{\rho^2}{\rho_1^2}G_1 G^{-1}G_1G^{-1}G_1  - \ldots\right]  \label{GBGexp} 
\end{alignat}

Since both $G$ and $G_1$ are self-adjoint, it is now evident that $GB^{-1}G_1$ is self-adjoint too. The substitution of \eqref{DN_G13} and \eqref{DN_G15} in \eqref{GBGexp} gives 

\begin{alignat}{2}
\nonumber
GB^{-1}G_1&= \delta^2\frac{h_1}{\rho_1} D^2 - \delta^3 \frac{1}{\rho_1 }\left( D \eta D + i  \frac{\rho h_1^2}{\rho_1}D^3 \mathcal{T}_h \right)  \\
 &    + \delta^4  \frac{\rho h_1}{\rho_1^2}\left(i D \eta D^2\mathcal{T}_h + i \mathcal{T}_h  D^2 \eta D \right) - \delta^4 \frac{h_1^3}{\rho_1} \left( \frac{1}{3}  + \frac{\rho^2}{\rho_1^2} \mathcal{T}_h^2\right) D^4  + \mathcal{O}
(\delta^5)  \label{GBG1exp} 
\end{alignat}
where the notation $\mathcal{T}_h:=-i\coth(hD)=(i\mathfrak{t}_h)^{-1}$ is introduced, more details are given in the Appendix \ref{ssec:A1}.  We need also 
\begin{align}
B^{-1}&=  \frac{1}{\rho_1}  \left[ G^{-1} -\frac{\rho}{\rho_1} G^{-1}G_1 G^{-1 } + \frac{\rho^2}{\rho_1^2}G^{-1}G_1 G^{-1}G_1G^{-1}  - \ldots\right] \nonumber \\
& =    \delta^{-1}\frac{1}{\rho_1}D^{-1}\mathfrak{t}_h^{-1}  + \mathcal{O}(1),  \label{Bm1exp} 
\end{align}

\begin{align}
B^{-1}G&=  \frac{1}{\rho_1}  \left[ 1 -\frac{\rho}{\rho_1} G^{-1}G_1  + \frac{\rho^2}{\rho_1^2}G^{-1}G_1 G^{-1}G_1  - \ldots\right] \nonumber \\
&= \frac{1}{\rho_1}  \left( 1 -   \delta \frac{\rho_1}{\rho_1}i\mathcal{T}_h D \right) + \mathcal{O} (\delta^2).   \label{Bm1Gexp} 
\end{align}
The quantity $\eta \eta_x= i h_1^3(\eta/h_1)\mathcal{D}_1(\eta/h_1)\sim \delta^3.$ The contribution of the integral density $ \eta \eta_x    B^{-1}  \eta \eta _x   $  in the Hamiltonian is therefore of order $\delta^5$. Recall that $dx \sim 1/\delta$. Hence, by keeping terms up to $\delta^3$ in \eqref{Main_approx} we have 
\begin{align}
\label{H_a}
H(\eta,\mathfrak{u})=&\delta \frac{h_1}{2\rho_1}\int_{\mathbb{R}} \mathfrak{u}^2 \,dx+\delta \frac{A}{2}\int_{\mathbb{R}}\eta^2 \, dx    +\delta \kappa  \int_{\mathbb{R}} \eta \mathfrak{u} \, dx  \nonumber \\  &-\delta^2 \frac{1}{2\rho_1} \int_{\mathbb{R}}  \eta \mathfrak{u} ^2   \,dx - \delta^2 \frac{h_1^2\rho}{2\rho_1^2} \int_{\mathbb{R}} \mathfrak{u} \mathcal{T}_h \mathfrak{u}_x   \,dx    + \delta^2 \frac{\gamma_1}{2}    \int_{\mathbb{R}} \eta^2 \mathfrak{u} \,dx \nonumber \\ &+ \delta^2 \frac{\rho \gamma^2-\rho_1\gamma_1^2}{6}    \int_{\mathbb{R}} \eta^3  \,dx 
    -\delta^3 \frac{h_1^3}{2\rho_1}\int_{\mathbb{R}} \mathfrak{u}_x\left(\frac{1}{3} +\frac{\rho^2}{\rho_1^2}\mathcal{T}_h^2  \right) \mathfrak{u}_x  \,  dx \nonumber \\ &+ \delta^3 \frac{h_1\rho}{\rho_1^2} \int_{\mathbb{R}} \eta \mathfrak{u} \mathcal{T}_h \mathfrak{u}_x   \,dx +\delta^3 \frac{(\gamma-\gamma_1)h_1\rho}{2\rho_1} \int_{\mathbb{R}} \eta ^2 \mathcal{T}_h \mathfrak{u}_x   \,dx  .
\end{align} 
where the constant $A=g(\rho-\rho_1)+\kappa(\rho \gamma - \rho_1 \gamma_1).$
We notice that $H$ is of order $\delta$. This gives the proper scaling of $\partial_t$ which should be also of order $\delta$, same as the order of $\partial_x$. The variation $\delta \mathfrak{u}$ bears a scale factor $\delta$ as well.  The equations \eqref{H_u_eta} with the scaling written explicitly therefore are

\begin{equation} \label{H_u_eta_scale}
\eta_t +(\delta )^{-1}\Big(\frac{\delta H}{\delta \mathfrak{u}}\Big)_x =0 \quad \text{and} \quad \mathfrak{u}_t+\Gamma\eta_t +(\delta )^{-1}\Big(\frac{\delta H}{\delta \eta}\Big)_x =0
\end{equation}
producing the coupled system 

\begin{align}
\eta_t + \kappa\eta_x &+\frac{h_1}{\rho_1}\mathfrak{u}_x-\delta\frac{1}{\rho_1}(\eta\mathfrak{u})_x
-\delta \frac{\rho h_1^2}{\rho_1^2} \mathcal{T}_h  \mathfrak{u}_{xx}
+\delta \gamma_1\eta\eta_x  +\delta^2\frac{h_1^3}{\rho_1} \Big(\frac{1}{3} +\frac{\rho^2}{\rho_1^2}\mathcal{T}_h^2  \Big) \mathfrak{u}_{xxx} \nonumber \\&+\delta^2\frac{\rho h_1}{\rho_1^2} \left( (\eta \mathcal{T}_h \mathfrak{u}_x)_x + \mathcal{T}_h (\eta \mathfrak{u})_{xx}\right) +\delta^2\frac{\rho h_1(\gamma-\gamma_1)}{2\rho_1} \mathcal{T}_h (\eta^2)_{xx}=0 \label{ILW_Eq1}\\
\label{ILW_Eq2}
 \mathfrak{u}_t  + \kappa\mathfrak{u}_x &+\Gamma\eta_t + A\eta_x -\delta\frac{1}{\rho_1}\mathfrak{u}\mathfrak{u}_x
+\delta \gamma_1 (\eta\mathfrak{u})_x +\delta(\rho\gamma^2-\rho_1\gamma_1^2) \eta\eta_x \nonumber \\ &+ \delta^2 \frac{\rho h_1}{\rho_1^2}(\mathfrak{u} \mathcal{T}_h \mathfrak{u}_x)_x  + \delta^2 \frac{\rho h_1(\gamma -\gamma_1)}{\rho_1}
(\eta \mathcal{T}_h \mathfrak{u}_x)_x =0.
\end{align}

These equations can be viewed as a generalisation of the irrotational case ($\Gamma=\gamma_1=\gamma=0,$  $\kappa=0$) derived in \cite{CraigGuyenneKalisch}.

\section{The intermediate long wave equation (ILWE)} \label{sec:6}

 Neglecting the terms of order $\delta^2$ in \eqref{ILW_Eq1}, \eqref{ILW_Eq2} we obtain
\begin{align}
\label{ILW_1}
&\eta_t + \kappa\eta_x +\frac{h_1}{\rho_1}\mathfrak{u}_x-\delta\frac{1}{\rho_1}(\eta\mathfrak{u})_x
-\delta \frac{\rho h_1^2}{\rho_1^2} \mathcal{T}_h  \mathfrak{u}_{xx}
+\delta \gamma_1\eta\eta_x
=0 \\
\label{ILW_2}
 &\mathfrak{u}_t  + \kappa\mathfrak{u}_x +\Gamma\eta_t + A\eta_x -\delta\frac{1}{\rho_1}\mathfrak{u}\mathfrak{u}_x
+\delta \gamma_1 (\eta\mathfrak{u})_x +\delta(\rho\gamma^2-\rho_1\gamma_1^2) \eta\eta_x
=0.
\end{align}

We can perform a Galilean transformation of coordinates  
\begin{equation} \label{GS}
X=x-\kappa t, \quad T=t, \quad \partial_X=\partial_x, \quad D\rightarrow-i\partial_X \quad   \mbox{ and } \quad \partial_T=\partial_t+\kappa \partial_x
\end{equation}
and taking into account that for the typical values of $\kappa$ of several m/s,  $g\gg 2\omega\kappa$ the equations of motion can be written as

\begin{align}
\label{ILW_10}
&\eta_T  +\frac{h_1}{\rho_1}\mathfrak{u}_X-\delta\frac{1}{\rho_1}(\eta\mathfrak{u})_X
-\delta \frac{\rho h_1^2}{\rho_1^2} \mathcal{T}_h  \mathfrak{u}_{XX}
+\delta \gamma_1\eta\eta_X
=0 \\
\label{ILW_20}
 &\mathfrak{u}_T  +\Gamma \eta_T    + (A-\Gamma \kappa)\eta_X -\delta\frac{1}{\rho_1}\mathfrak{u}\mathfrak{u}_X
+\delta \gamma_1 (\eta\mathfrak{u})_X +\delta(\rho\gamma^2-\rho_1\gamma_1^2) \eta\eta_X
=0,
\end{align}

$A-\Gamma \kappa =g(\rho-\rho_1)+\kappa(\rho \gamma - \rho_1 \gamma_1) -\kappa(\rho\gamma-\rho_1\gamma_1)+2\kappa\omega\big(\rho-\rho_1\big)
= (g-2\kappa \omega)(\rho-\rho_1)\approx g(\rho-\rho_1), $ finally

\begin{align}
\label{ILW_101}
&\eta_T  +\frac{h_1}{\rho_1}\mathfrak{u}_X-\delta\frac{1}{\rho_1}(\eta\mathfrak{u})_X
-\delta \frac{\rho h_1^2}{\rho_1^2} \mathcal{T}_h  \mathfrak{u}_{XX}
+\delta \gamma_1\eta\eta_X
=0 \\
\label{ILW_202}
 &\mathfrak{u}_T  +\Gamma \eta_T    + g(\rho-\rho_1)\eta_X -\delta\frac{1}{\rho_1}\mathfrak{u}\mathfrak{u}_X
+\delta \gamma_1 (\eta\mathfrak{u})_X +\delta(\rho\gamma^2-\rho_1\gamma_1^2) \eta\eta_X
=0,
\end{align}

A two-dimensional version of \eqref{ILW_101} -- \eqref{ILW_202} in the irrotational case ($\Gamma=\gamma_1=\gamma=0$) is obtained in \cite{BonnaLannesSaut}. The leading order terms (i.e. neglecting the terms with $\delta$ above) produce a system of linear equations with constant coefficients from where the speed(s) of the travelling waves (in the leading order) could be determined:

 \begin{equation} \label{c4BO}
c=-\frac{h_1}{2\rho_1}\Gamma \pm \sqrt{\frac{h_1^2}{4\rho_1^2}\Gamma^2+\frac{h_1}{\rho_1}g(\rho-\rho_1)}. \end{equation} The plus sign is for the right-running waves and the minus sign is for the left-running waves. These speeds coincide with the speeds in the case of infinitely deep lower layer \cite{CI19}, indeed, the $h$-dependence comes only from the term $\mathcal{T}_h$ which is of order $\delta$.

In what follows, $c$ could be either of the two solutions of the dispersion equation, the choice of solution determines if the results are relevant for the left or the right-running waves.  

For the travelling wave, which depends on the characteristic variable $X-c T,$ we also have $\mathfrak{u}=\frac{\rho_1}{h_1}c \eta$ and in order to obtain a single nonlinear equation for $\eta$ we expect a relation which involves terms of order $\delta$ as well. In other words, we consider an expansion of the form
\begin{equation}
        \mathfrak{u}=\frac{\rho_1}{h_1}c \eta+\delta\alpha\eta^2+\delta\beta \mathcal{T}_h\eta_X,\label{JT}
\end{equation} for some yet undetermined constants $\alpha$ and $\beta$. This type of relation is known also as the Johnson transformation. The substitution of $\mathfrak{u}$ from \eqref{JT} in  \eqref{ILW_10} and  \eqref{ILW_20} when keeping only the terms up to order $\delta$ leads to two equations for $\eta$ and therefore these two equations must coincide. This leads to equality of the coefficients in front of the terms of the same type, which further allows to determine the previously unknown 
\begin{equation} \label{alpha}
\alpha
=
 \frac{\rho_1 (\rho_1 c^2
+2 h_1 \Gamma c
- \gamma_1 h_1^2 \Gamma
+\rho_1 \gamma_1 h_1 c
+h_1^2 ) }{2h_1^2 \big(2\rho_1c+h_1\Gamma  \big)}
\end{equation} and \begin{equation} \label{beta}
\beta
 =\frac{\rho(\rho_1c^2+  h_1 \Gamma c)}{2\rho_1c+h_1 \Gamma}.
\end{equation}

The equation for $\eta$ is 
\begin{equation} \label{ILW}
\eta_T+c\eta_X -\delta\frac{\rho h_1c^2}{2\rho_1 c+h_1 \Gamma }\mathcal{T}_h\eta_{XX}
+\delta \frac{-3\rho_1c^2
+3\rho_1 \gamma_1 h_1c
+ h_1^2 (\rho\gamma^2-\rho_1\gamma_1^2) }{h_1(2\rho_1c+h_1\Gamma)}\eta\eta_X=0.
\end{equation}
The obtained equation is known as the Intermediate Long Wave Equation (ILWE) introduced in \cite{J,K}.  It is an integrable equation. The soliton theory for ILWE has been developed in a number of works of which we mention \cite{ChLee,JoEg,Kod,Ma2,SaAbKo}. 

The ILWE in the irrotational case ($\gamma=\gamma_1=\omega=0,$  $\kappa=\Gamma=0$) becomes\footnote{Note that in this case $\kappa=0$ and hence $(X,T)\equiv (x,t)$.}

\begin{equation}
\eta_t+c\eta_x -\delta\frac{\rho h_1 c}{2\rho_1 } \mathcal{T}_h\eta_{xx}
-\delta \frac{3c}{2h_1} \eta\eta_x=0,
\end{equation} where, from \eqref{c4BO}
\begin{equation}
c=\pm \sqrt{\frac{h_1}{\rho_1}g(\rho-\rho_1)}. \notag
\end{equation} 

Let us write the ILWE in the form
\begin{equation} \label{ILW1}
        \eta_T +c\eta_X+\delta\mathcal{A}\eta \eta_X -\delta\mathcal{B}\mathcal{T}_h\eta_{XX} =0
\end{equation} 
where \begin{equation} \label{A} 
\mathcal{A}:=  \frac{-3\rho_1c^2
+3\rho_1 \gamma_1 h_1c
+ h_1^2 (\rho\gamma^2-\rho_1\gamma_1^2) }{h_1(2\rho_1c+h_1\Gamma)}, \qquad     \mathcal{B}:= \frac{\rho h_1c^2}{2\rho_1 c+h_1 \Gamma } .\end{equation}
    
Details about the operator $\mathcal{T}_h$ and the integrability of the ILWE are given in the Appendix \ref{ssec:A1} and \ref{ssec:A2} respectively. Here we only mention that the one-soliton solution of \eqref{ILW1} has the form (cf. \eqref{sU1}, Appendix \ref{ssec:A2} )   
   
 \begin{align} \label{ILWsol}\eta(X,T)= \frac{2\mathcal{B}}{\mathcal{A}}\cdot \frac{k_0 \sin(k_0 h)}{\cos(k_0 h)+\cosh [k_0 (X-X_0-(c-\delta \mathcal{B}k_0 \cot(k_0 h)) T)]},& \\
 0<k_0<  \frac{2\pi}{h}.& \nonumber 
   \end{align}  
In the above formula $X_0$ and $k_0$ are the soliton parameters, i.e. arbitrary constants within their range of allowed values. $X_0$ is the initial position of the crest of the soliton and $k_0$ is related to its amplitude.  The wavespeed of the soliton is  $c - \delta \mathcal{B}k_0 \cot(k_0 h) $ and the correction of order $\delta$ depends on the coefficient  $\mathcal{B}$ and the dispersion law related to the  dispersive term and also on the parameter $k_0$ which is related to the discrete eigenvalue of the spectral problem \eqref{1}; in the 1-soliton case there is only one discrete eigenvalue. We observe that both the amplitude of the soliton and its speed are related through $k_0$. Another feature of this solution is the fact that the function $\cot$ is unbounded. The physical relevance of the solution however requires that the choice of $k_0$ should be such that the quantity $k_0\cot(k_0 h)$ is of order 1. As it will be shown in the next section, there is no such anomaly for the related Benjamin-Ono equation, and the limiting procedure requires a special choice of the parameter $k_0$.      

\section{Connection to the Benjamin-Ono equation}

The BO model of waves in the presence of uniformly-sheared currents has been derived previously in \cite{CI19}. The mathematical facts about the BO equation are given in Appendix \ref{ssec:A2} for convenience. In this section we demonstrate that the BO equation could be obtained as a special kind of a long-wave limit from the ILWE.

In the limit $h\to \infty$ which corresponds to an infinitely deep lower layer we have $$\mathcal{T}_h= -i\coth(hD) \to -i \,\text{sign}(D), \qquad \mathcal{T}_h\partial_X \to |D|,$$  and the equation \eqref{ILW} becomes the well known Benjamin-Ono (BO) equation \cite{BO1,BO2}, see also \cite{CI19},
\begin{equation}
        \eta_T +c\eta_X+\delta\mathcal{A}\eta \eta_X -\delta\mathcal{B}|D|\eta_{X} =0
\end{equation} 
Like the ILWE, the BO is an integrable equations whose solutions can be obtained by the Inverse Scattering method \cite{FA,KM98,Ma1}.

The limit to the one soliton solution of the BO equation (from \eqref{ILWsol} with $h \to \infty$ but $k_0 h$ finite, $ k_0 h = \pi - k_0/q$ where $q$ is a constant) is described in the Appendix \ref{ssec:A2} cf. \eqref{BOLim} and can be written in the form
\begin{equation} \eta(X,T)=\frac{\eta_0}{1+\left(\frac{\eta_0 \mathcal{A}}{4\mathcal{B}}\right)^2[X-X_0-(c+\frac{1}{4}\delta  \mathcal{A}\eta_0) T]^2} \end{equation} where the constants are the initial position $X_0$ of the soliton and its amplitude $\eta_0.$  The relation to the constant $q$ (and hence $k_0$) is $$\eta_0=4\mathcal{B}q/\mathcal{A}=4\mathcal{B}k_0/[\mathcal{A}(\pi -k_0 h)].$$

\section{Connection to the KdV equation}

The KdV model of waves in the presence of uniformly-sheared currents has been derived previously in \cite{CompelliIvanov2}. The special situation of KdV with $h_1/h \sim \delta$ has been provided in  Appendix \ref{ssec:KdV} for convenience. The provided analysis shows that it could be obtained as a special kind of a long-wave limit from the ILWE.  
  
The KdV limit could be obtained assuming $hk\ll 1$ or $|hD| \ll1.$ The operator 
$$\mathcal{T}_h =-i \coth(hD) \approx -i\left(\frac{1}{hD} +\frac{1}{3} hD \right)=\frac{1}{h} \partial_X^{-1}-\frac{1}{3}h\partial_X.$$
Then from \eqref{ILW1} we have
\begin{equation} \label{KdV-1}
        \eta_T +\left(c-\delta \frac{\rho c^2}{2\rho_1 c +h_1 \Gamma} \frac{h_1}{h}\right)\eta_X+\delta\mathcal{A}\eta \eta_X +\delta\frac{h\mathcal{B}}{3} \eta_{XXX} =0.
\end{equation} Noting that $h_1/h \simeq \delta \ll1$ the correction to $c$ in the second term is of order $\delta^2$ and should be neglected. Thus we obtain the KdV equation in the form 
\begin{equation} \label{KdV-2}
        \eta_T + c\eta_X+\delta\mathcal{A}\eta \eta_X +\delta\frac{h\mathcal{B}}{3} \eta_{XXX} =0.
\end{equation} which coincides with \eqref{KdVbb}, Appendix \ref{ssec:KdV} since in the ILWE setup $\delta \simeq \varepsilon $ and $$\mathcal{B}_1= \frac{h\mathcal{B}}{3} =\frac{c^2 \rho h h_1}{3(2c\rho_1+\Gamma h_1) }.$$

The one-soliton solution of \eqref{KdV-2} can also be obtained from \eqref{ILWsol}. For $k_0 h \ll 1$ we have $\sin(k_0 h) \approx k_0 h,$  $\cos(k_0 h) \approx  1 ,$  and also the identity  
\begin{align} \nonumber  1+\cosh Z &= 2 \cosh ^2 (Z/2) ,\\  \nonumber  \mathcal{B}k_0\cot(k_0 h) &\approx  \frac{3 \mathcal{B}_1}{h}k_0 \left(\frac{1}{k_0 h} - \frac{k_0 h}{3} \right)= \frac{3 \mathcal{B}_1}{h^2}-\mathcal{B}_1 k_0^2 \\ \nonumber
& =  \frac{\rho c^2  }{2c\rho_1+\Gamma h_1 }\cdot \frac{h_1}{h} -\mathcal{B}_1 k_0^2   \end{align}  The first term $$\frac{c^2 \rho }{2c\rho_1+\Gamma h_1 }\cdot \frac{h_1}{h}=\frac{c^2 \rho }{2c\rho_1+\Gamma h_1 }\cdot \delta \ll1 $$ does not depend on $k_0$ and represents the small $\mathcal{O}(\delta^2)$ correction to the constant wave speed $c$. The second term is proportional to $k_0^2$, and the approximation leads to 
 \begin{align} \eta(X,T)&= \frac{6\mathcal{B}_1}{\mathcal{A}}\cdot \frac{k_0^2  }{1+\cosh [k_0 (X-X_0-(c+\delta \mathcal{B}_1 k_0^2 ) T)]} \nonumber \\ &=  \frac{3\mathcal{B}_1}{\mathcal{A}}\cdot \frac{k_0^2  }{\cosh^2 [\frac{k_0}{2} (X-X_0-(c+\delta \mathcal{B}_1k_0^2 ) T)]}    .   \label{KdV-lim-sol} \end{align} Introducing a new constant  $K={k_0}/{2} ,$   we finally have 
    \begin{equation} \label{KdV-lim-final}\eta(X,T)= \frac{12\mathcal{B}_1}{\mathcal{A}}\cdot \frac{K ^2  }{\cosh^2 [K (X-X_0-(c+ 4\delta K^2 \mathcal{B}_1  ) T)]}  \end{equation}
which coincides with the KdV one-soliton solution \eqref{KdV1sol}, Appendix \ref{ssec:KdV}.

  \section{Conclusions}
  
  We have derived the integrable ILWE for the situation of solitary waves on the interface of two fluids with constant vorticities, modeling equatorial internal waves interacting with uniformly sheared currents. The surface waves which are usually of much smaller amplitude are neglected, so a ``rigid lid'' approximation for the upper fluid is assumed. The ILWE is an integrable model and the inverse scattering method or other methods like Darboux transforms and Hirota's method allow for the derivation of explicit multisoliton solutions. The one-soliton solution for example is \eqref{ILWsol}.  The two important integrable limits leading to the BO and KdV equations (for specific choice of the physical quantities) are explained in details. In addition these two limits have been applied to the ILWE one-soliton solution, and the BO and KdV one-soliton solutions have been recovered. The limits of course exist for the multisoliton ILWE solutions as well and in principle for all types of solutions.
It has to be noted that the ILWE, BO and KdV correspond to different scales of the physical quantities, as it could be seen from Table \ref{tab:table1}.
The BO limit corresponds to a limit to an infinitely deep lower layer. The limit from ILWE to KdV gives a KdV equation in its particular form \eqref{KdV-2}. Schematically the relations between the ILWE and the KdV equations \eqref{KdVa}, \eqref{KdVbb} and \eqref{KdV-2} could be represented as follows: 
$$\text{KdV}\, (\ref{KdVa})  \xrightarrow[]{(h_1/h) = \varepsilon\ll1}\text{KdV}\, (\ref{KdVbb}) \xrightarrow[]{ \delta:=\varepsilon } \fbox{\text{KdV}\, (\ref{KdV-2})} \xleftarrow[]{hk\ll1} \text{ILWE} $$

This shows that the ILWE should be used as a ``master'' equation with some caution. An interesting aspect for further studies is the development of theoretical models for internal waves with currents over variable bottom, extending the results from \cite{Na1,Na2}.

\begin{center}

\begin{table}[]
\begin{tabular}{|l|l|l|l|}
\hline
 &\text{ILWE}  & \text{BO}  & \text{KdV}\quad (\ref{KdVa})   \\ \hline
$\mathcal{O}(h_1/h) $ & $\delta$ & 0 & 1  \\ \hline
$\mathcal{O}(\eta/h_1)$ & $\delta$ & $\delta$ & $\delta^2$      \\ \hline
$\mathcal{O}(h_1k)$& $\delta$ & $\delta$ & $\delta$    \\ \hline
$\mathcal{O}(hk)$ &     1 & $\infty$ & $\delta$  \\ \hline
\end{tabular}
\caption{\label{tab:table1} The scales of the three approximation models. }
\end{table}

\end{center}

\appendix

\section{Appendix}\label{sec:A}
\subsection{The operator $\mathcal{T}_h$}\label{ssec:A1}
The action of operators like $\mathcal{T}_h=-i\coth(hD)$ is defined with the help of Fourier transforms $\mathcal{F}$, so let us introduce   $$\hat{v}(k):=\mathcal{F}^{-1}\{v(x)\}(k),\qquad v(x)=\mathcal{F}\{\hat{v}(k)\}(x).$$
Then $$\mathcal{T}_h v(x):=-i\mathcal{F}\{\coth(hk)\hat{v}(k)\}(x)$$ and furthermore \begin{align}\mathcal{T}_h v(x)&=-i\frac{1}{\sqrt{2\pi}}\int e^{ikx}\coth(hk)\hat{v}(k)  dk \nonumber \\
&=\frac{-i}{\sqrt{2\pi}}\int e^{ikx}\coth(hk)\left(\frac{1}{\sqrt{2\pi}}\int e^{-ikx'} v(x')dx' \right)  dk \nonumber \end{align}
   By changing the order of integration and using some appropriate integrals from \cite{RG} for the integration over $dk$  we obtain finally
$$\mathcal{T}_h v(x)=-\frac{1}{2h} \text{P.V.} \int_{-\infty}^{\infty} \coth \frac{\pi(x-x')}{2h} v(x') dx'. $$
When $h\to \infty,$  $$\frac{1}{2h}\coth\frac{\pi(x-x')}{2h}\to \frac{1}{2h}\cdot \frac{2h}{\pi(x-x')}=\frac{1}{\pi(x-x')}$$ and $\mathcal{T}_h$ becomes the Hilbert transform, $\mathcal{H}$ 
$$ \mathcal{T}_h v(x)\to \mathcal{H}\{v\} (x) := \mathrm{P.V.}\frac{1}{\pi}\int_{-\infty}^{\infty}\frac{v(x')dx'}{x-x'}.$$

\subsection{The integrability of the ILWE and its limit to the BO equation}
\label{ssec:A2}

Let us consider the class of complex valued analytic functions, defined in the complex plane $z=x+iy,$ $x,y \in \mathbb{R},$ which are $2h$-periodic with respect to the $y$ variable, i.e. $\Psi(x+i0)=\Psi(x+i2h).$ Let us consider the fundamental domain of this class, $0\le y< 2h$ and let us define 
$$\Psi^+(x)=\Psi(x+i0^+), \qquad \Psi^-(x)=\Psi(x+i2h - i0^+).$$ The Lax representation for the ILWE written in the form 
\begin{equation} \label{U} U_t -CU_x -2UU_x + \mathcal{T}_h U_{xx}=0 \end{equation}
($a$ is an arbitrary real constant) involves the compatibility of a Riemann-Hilbert Problem (RHP) and another linear problem for $\Psi(z,t,\lambda), $ ($\lambda$ is a spectral parameter)
\begin{align}
&i\Psi_x^+ + U \Psi^+=\lambda \Psi^- ,\label{1}\\
& i \Psi_t^{\pm}=\alpha_0^{\pm}\Psi^{\pm}+\alpha_1^{\pm}\Psi^{\pm}_x +\Psi^{\pm}_{xx} \label{2} 
\end{align}  
for some functions $\alpha_n^{\pm}(x,t), $ $n=0,1$ which are limits of some periodic functions  $\alpha_n^{\pm} (z,t), $ of the same class.
The second equation \eqref{2} multiplied by $\lambda$ and transformed with \eqref{1} gives
\begin{align}
i \lambda\Psi_t^{-}=&\alpha_0^{-}(\lambda \Psi^{-})+\alpha_1^{-}(\lambda\Psi^{-})_x +(\lambda\Psi^{-})_{xx}  \nonumber \\  
 = & i \Psi^{+}_{xxx}+(U+i\alpha_1^-)\Psi^+_{xx}+(2U_x+U\alpha_1^- +i\alpha_0^-)\Psi^+_{x}\nonumber
\\
& \phantom{*****}+(U_{xx}+\alpha_1^-U_x +U\alpha_0^-)\Psi^+ \label{32}
\end{align} 
On the other hand, differentiating the first equation \eqref{1} with respect to $t$ and then transforming with \eqref{2} we have
\begin{align}
i (\lambda\Psi^{-})_t=&i(i \Psi^{+}_t)_x +iU_t \Psi^+ +U(i\Psi^{+})_{t}  \nonumber \\  
 =& i \Psi^{+}_{xxx}+(U+i\alpha_1^+)\Psi^+_{xx}+(i\alpha^+_{1,x} +i\alpha_0^+ + U \alpha_1^+)\Psi^+_{x}   \nonumber \\ &  \phantom{*****}+  (i\alpha_{0,x}^+ + i U_t+\alpha_0^+U )\Psi^+ \label{33}
\end{align} 
From \eqref{32}, \eqref{33} we have $\alpha_1^+=\alpha_1^-$ so $\alpha_1^{\pm}=iC=\text{const};$ $U$ satisfying \begin{equation} \label{Ueq} U_t-CU_x-2UU_x +iU_{xx}+\alpha_{0,x}^+=0 \end{equation} with $\alpha_0(z)$ solution of the RH Problem \begin{equation} \label{RHP}\alpha_0^+-\alpha_0^-=-2iU_x\equiv v(x).\end{equation}  If ($\sigma$ is a constant)
$$\alpha_0^+(x)=\frac{1}{\sqrt{2\pi}}\int e^{ikx}\hat{\alpha}_0(k) dk + \sigma $$ then $$ \alpha_0^-(x)=\frac{1}{\sqrt{2\pi}}\int e^{ikx}\hat{\alpha}_0(k)e^{-2hk} dk+\sigma ,$$ and from \eqref{RHP}  \begin{equation} \label{v} \hat{\alpha}_0(k)[1-e^{-2hk}]=\hat{v}(k)\end{equation}
and hence 
\begin{align}\alpha_0^+(x)+ \alpha_0^-(x)&=\frac{1}{\sqrt{2\pi}}\int e^{ikx}\hat{\alpha}_0(k)[1+e^{-2hk}] dk + 2\sigma \nonumber \\ &=\frac{1}{\sqrt{2\pi}}\int e^{ikx}\hat{v}(k)\coth(hk) dk + 2\sigma=i\mathcal{T}_h v(x)+2\sigma, \nonumber \end{align}
$\alpha_0^+(x)+ \alpha_0^-(x)=i\mathcal{T}_h(-2iU_x)+2\sigma=2\mathcal{T}_h U_x+2\sigma$ and using the RHP \eqref{RHP} we have $\alpha_0^{+}=\mathcal{T}_h U_x-iU_x+\sigma.$ The substitution in \eqref{Ueq} gives \eqref{U}. Note that if the jump condition in the RHP is $v\equiv 0$ then due to \eqref{v} the corresponding analytic function can only be a constant.
The inverse scattering method for the ILWE is developed in \cite{Kod}. Equation \eqref{U} after the change $U\to -U$ and $C\to -C$ can be written as
\begin{equation} \label{U1} U_t +CU_x +2UU_x + \mathcal{T}_h U_{xx}=0 \end{equation}
with one soliton solution \begin{equation} \label{sU1} U(x,t)= \frac{k_0 \sin (k_0 h)}{\cos (k_0 h ) + \cosh [k_0(x-x_0 -(C-k_0 \cot(k_0 h))t)]} \end{equation} where $k_0$ and $x_0$ are constants.

The connection between the inverse scattering methods for the ILWE and the BO equations is studied in \cite{Sant}. The limit $h\to \infty$ can be performed only together with $k_0 \to 0$ and keeping $k_0 h $ finite such that $$k_0 h = \pi - \frac{k_0}{q}$$ where $q$ is a constant. Then $\sin(k_0 h)=\sin(\frac{k_0}{q}) \approx \frac{k_0}{q}, $  $\cos(k_0 h) = -\cos\frac{k_0}{q} \approx -1+\frac{k_0^2}{2q^2}; $ $\cosh Z \approx 1+\frac{1}{2}Z^2.$ This way we obtain the one soliton solution of the BO equation \begin{equation} \label{BO1} U_t +CU_x +2UU_x + |D| U_{x}=0 \end{equation} in the form
\begin{equation} \label{BOLim} U(x,t)=\frac{2q}{1+q^2[x-x_0-(C+q)t]^2} \end{equation}
where now the constants are $x_0$ and $q$.

\subsection{The KdV equation for internal waves with currents}\label{ssec:KdV}

The propagation of internal waves with currents in the KdV regime $\varepsilon\simeq \delta^2 \ll 1$ has been derived in \cite{CompelliIvanov2}. Without the assumption that $h_1/h $ is small, the equation for the elevation $\eta(X,T)$ is  
\begin{equation} \label{KdVa} 
\eta_T+c\eta_X+\varepsilon\frac{c^2 \alpha_2}{\alpha_1(2c+\Gamma \alpha_1) }\eta_{XXX}
+\varepsilon\frac{3c^2\alpha_3+3c\alpha_1\alpha_4+\alpha_1^2\alpha_5 }{\alpha_1(2c+\Gamma \alpha_1) }\eta \eta_X=0.
\end{equation} where 

\begin{equation} \begin{split} \alpha_1&=\frac{h h_1}{\rho_1 h+\rho h_1}, \qquad \alpha_2= \frac{h^2h_1^2(\rho h+\rho_1 h_1)}{3(\rho_1 h+\rho h_1)^2}, \qquad \alpha_3= \frac{\rho h_1^2-\rho_1 h^2}{(\rho_1 h+\rho h_1)^2}, \\
\alpha_4&=\frac{\gamma_1\rho_1 h+ \gamma \rho h_1}{\rho_1 h+\rho h_1}, \qquad  \alpha_5=\rho\gamma^2-\rho_1 \gamma_1^2, \\
  c&= -\frac{\alpha_1 \Gamma}{2} \pm \sqrt{\frac{\alpha_1 ^2 \Gamma^2}{4}+ \alpha_1(\rho-\rho_1)g}.
\end{split}\end{equation}
With the further assumption $\frac{h_1}{h}=\varepsilon \ll1$ we have the following values of the constants:

\begin{equation} \begin{split} \alpha_1& \approx\frac{ h_1}{\rho_1 }, \qquad \alpha_2 \approx \frac{\rho h  h_1 ^2}{3\rho_1 ^2}, \qquad \alpha_3 \approx - \frac{1}{\rho _1}, \\
\alpha_4& \approx \gamma _1 , \quad  \alpha_5=\rho\gamma^2-\rho_1 \gamma_1^2, \quad c= -\frac{h_1 \Gamma}{2 \rho_1} \pm \sqrt{\frac{h_1 ^2 \Gamma^2}{4\rho_1^2}+ \frac{(\rho-\rho_1)g h_1}{\rho_1}    }.
\end{split}\end{equation}

We notice that $c$ coincides with \eqref{c4BO}, the wave speed of the ILW equation. The KdV equation acquires the form 

\begin{equation} \label{KdVb} 
\eta_T+c\eta_X+\varepsilon\frac{c^2 \rho h h_1}{3(2c\rho_1+\Gamma h_1) }\eta_{XXX}
+\varepsilon\frac{-3\rho_1 c^2 +3\rho_1 h_1\gamma_1 c+h_1^2 (\rho\gamma^2-\rho_1 \gamma_1^2) }{h_1(2c\rho_1+\Gamma h_1) }\eta \eta_X=0,
\end{equation} 
or

\begin{equation} \label{KdVbb} 
\eta_T+c\eta_X+\varepsilon \mathcal{B}_1  \eta_{XXX}
+\varepsilon \mathcal{A}\eta \eta_X=0,
\end{equation} 
with $$\mathcal{B}_1=\frac{c^2 \rho h h_1}{3(2c\rho_1+\Gamma h_1) }, \qquad \mathcal{A}=\frac{-3\rho_1 c^2 +3\rho_1 h_1\gamma_1 c+h_1^2 (\rho\gamma^2-\rho_1 \gamma_1^2) }{h_1(2c\rho_1+\Gamma h_1) },$$ moreover $\mathcal{A}$ coincides with the expression from \eqref{A} and $\mathcal{B}_1= h\mathcal{B}/3.$ The one-soliton solution is (see for example \cite{Johnson_Book} for the KdV solitons) 
\begin{equation} \label{KdV1sol}
\eta(X,T)=\frac{12\mathcal{B}_1}{\mathcal{A}}\frac{K^2}{\cosh^2[K(X-X_0-(c+\varepsilon 4K^2\mathcal{B}_1)T)]} 
\end{equation}
where $K$ and $X_0$ are constants.
%
%
%

\subsection*{Acknowledgments} One of the authors (RI) would like to thank the Erwin Schr\"odinger International Institute for Mathematics and Physics (ESI), Vienna (Austria)  for the opportunity to participate in the workshop {\it Mathematical Aspects of Geophysical Flows, } 20-24 January 2020 where a significant part of this work has been accomplished. The authors are thankful to all referees for their valuable comments and suggestions.



\end{document}